\documentclass[ALICE,manyauthors]{cernphprep}
\usepackage[comma,square,numbers,sort&compress]{natbib}
\usepackage{hyperref}
\usepackage{lineno}
\usepackage{xspace}
\usepackage{xcolor}
\usepackage{graphicx}
\usepackage[T1]{fontenc}
\usepackage{orcidlink}

\begin{document}
%%%%%%%%%%%%%%%%%%%%%%%%%%%%%%%%%%%%%%%%%%%%%%%%%%
% These are some new commands that may be useful 
% for paper writing in general. If other newcommands
% are needed for your specific paper, please feel 
% free to add here. 
%
% The currently available commands are organized in: 
% 1) Systems
% 2) Quantities
% 3) Energies and units
% 4) Detectors
% 5) particle species 
%%%%%%%%%%%%%%%%%%%%%%%%%%%%%%%%%%%%%%%%%%%%%%%%%%

% 1) SYSTEMS 
\newcommand{\pp}           {pp\xspace}
\newcommand{\ppbar}        {\mbox{$\mathrm {p\overline{p}}$}\xspace}
\newcommand{\XeXe}         {\mbox{Xe--Xe}\xspace}
\newcommand{\PbPb}         {\mbox{Pb--Pb}\xspace}
\newcommand{\pA}           {\mbox{pA}\xspace}
\newcommand{\pPb}          {\mbox{p--Pb}\xspace}
\newcommand{\AuAu}         {\mbox{Au--Au}\xspace}
\newcommand{\dAu}          {\mbox{d--Au}\xspace}

% 2) QUANTITIES 
\newcommand{\s}            {\ensuremath{\sqrt{s}}\xspace}
\newcommand{\snn}          {\ensuremath{\sqrt{s_{\mathrm{NN}}}}\xspace}
\newcommand{\pt}           {\ensuremath{p_{\rm T}}\xspace}
\newcommand{\meanpt}       {$\langle p_{\mathrm{T}}\rangle$\xspace}
\newcommand{\ycms}         {\ensuremath{y_{\rm CMS}}\xspace}
\newcommand{\ylab}         {\ensuremath{y_{\rm lab}}\xspace}
\newcommand{\etarange}[1]  {\mbox{$\left | \eta \right |~<~#1$}}
\newcommand{\yrange}[1]    {\mbox{$\left | y \right |~<~#1$}}
\newcommand{\dndy}         {\ensuremath{\mathrm{d}N_\mathrm{ch}/\mathrm{d}y}\xspace}
\newcommand{\dndeta}       {\ensuremath{\mathrm{d}N_\mathrm{ch}/\mathrm{d}\eta}\xspace}
\newcommand{\avdndeta}     {\ensuremath{\langle\dndeta\rangle}\xspace}
\newcommand{\dNdy}         {\ensuremath{\mathrm{d}N_\mathrm{ch}/\mathrm{d}y}\xspace}
\newcommand{\Npart}        {\ensuremath{N_\mathrm{part}}\xspace}
\newcommand{\Ncoll}        {\ensuremath{N_\mathrm{coll}}\xspace}
\newcommand{\dEdx}         {\ensuremath{\textrm{d}E/\textrm{d}x}\xspace}
\newcommand{\RpPb}         {\ensuremath{R_{\rm pPb}}\xspace}

% 3) ENERGIES, UNITS
\newcommand{\nineH}        {$\sqrt{s}~=~0.9$~Te\kern-.1emV\xspace}
\newcommand{\seven}        {$\sqrt{s}~=~7$~Te\kern-.1emV\xspace}
\newcommand{\twoH}         {$\sqrt{s}~=~0.2$~Te\kern-.1emV\xspace}
\newcommand{\twosevensix}  {$\sqrt{s}~=~2.76$~Te\kern-.1emV\xspace}
\newcommand{\five}         {$\sqrt{s}~=~5.02$~Te\kern-.1emV\xspace}
\newcommand{\twosevensixnn}{$\sqrt{s_{\mathrm{NN}}}~=~2.76$~Te\kern-.1emV\xspace}
\newcommand{\fivenn}       {$\sqrt{s_{\mathrm{NN}}}~=~5.02$~Te\kern-.1emV\xspace}
\newcommand{\LT}           {L{\'e}vy-Tsallis\xspace}
\newcommand{\GeVc}         {Ge\kern-.1emV/$c$\xspace}
\newcommand{\MeVc}         {Me\kern-.1emV/$c$\xspace}
\newcommand{\TeV}          {Te\kern-.1emV\xspace}
\newcommand{\GeV}          {Ge\kern-.1emV\xspace}
\newcommand{\MeV}          {Me\kern-.1emV\xspace}
\newcommand{\GeVmass}      {Ge\kern-.2emV/$c^2$\xspace}
\newcommand{\MeVmass}      {Me\kern-.2emV/$c^2$\xspace}
\newcommand{\lumi}         {\ensuremath{\mathcal{L}}\xspace}

% 4) DETECTORS 
\newcommand{\ITS}          {\rm{ITS}\xspace}
\newcommand{\TOF}          {\rm{TOF}\xspace}
\newcommand{\ZDC}          {\rm{ZDC}\xspace}
\newcommand{\ZDCs}         {\rm{ZDCs}\xspace}
\newcommand{\ZNA}          {\rm{ZNA}\xspace}
\newcommand{\ZNC}          {\rm{ZNC}\xspace}
\newcommand{\SPD}          {\rm{SPD}\xspace}
\newcommand{\SDD}          {\rm{SDD}\xspace}
\newcommand{\SSD}          {\rm{SSD}\xspace}
\newcommand{\TPC}          {\rm{TPC}\xspace}
\newcommand{\TRD}          {\rm{TRD}\xspace}
\newcommand{\VZERO}        {\rm{V0}\xspace}
\newcommand{\VZEROA}       {\rm{V0A}\xspace}
\newcommand{\VZEROC}       {\rm{V0C}\xspace}
\newcommand{\Vdecay} 	   {\ensuremath{V^{0}}\xspace}

% 4) PARTICLE SPECIES 
\newcommand{\hyp}         {${}^{3}_{\Lambda}\mathrm{H}\xspace}
\newcommand{\ee}           {\ensuremath{e^{+}e^{-}}} 
\newcommand{\pip}          {\ensuremath{\pi^{+}}\xspace}
\newcommand{\pim}          {\ensuremath{\pi^{-}}\xspace}
\newcommand{\kap}          {\ensuremath{\rm{K}^{+}}\xspace}
\newcommand{\kam}          {\ensuremath{\rm{K}^{-}}\xspace}
\newcommand{\pbar}         {\ensuremath{\rm\overline{p}}\xspace}
\newcommand{\kzero}        {\ensuremath{{\rm K}^{0}_{\rm{S}}}\xspace}
\newcommand{\lmb}          {\ensuremath{\Lambda}\xspace}
\newcommand{\almb}         {\ensuremath{\overline{\Lambda}}\xspace}
\newcommand{\Om}           {\ensuremath{\Omega^-}\xspace}
\newcommand{\Mo}           {\ensuremath{\overline{\Omega}^+}\xspace}
\newcommand{\X}            {\ensuremath{\Xi^-}\xspace}
\newcommand{\Ix}           {\ensuremath{\overline{\Xi}^+}\xspace}
\newcommand{\Xis}          {\ensuremath{\Xi^{\pm}}\xspace}
\newcommand{\Oms}          {\ensuremath{\Omega^{\pm}}\xspace}
\newcommand{\degree}       {\ensuremath{^{\rm o}}\xspace}

%%%%%%%%%%%%%%%%%%%%%%%%%%%%%%%%%%%%%%%%%%%%%%%%%%%%%%%%%%%%%%%%%%%%%%%%%%%%%%%%%%%%%%%%%%%%%%%%%%%%%%%%%%%%%%%%%%%%%%%%%%%%%%%%%%%%%%%%%%%%%%%%%%%%%
\begin{titlepage}
% the dates below correspond to CERN approval
% please don't touch: EB chairs will take care
\PHyear{2026}           % required, will be obtained from CERN
\PHnumber{079}        % required, will be obtained from CERN
\PHdate{13 March}  % required, will be obtained from CERN
%%%%%%%%%%%%%%%%%%%%%%%%%%%%%%%%%%%%%%%%%%%%%%%%%%%%%%%%%%%%%%%%%%%%%%%%%%%%%%%%%%%%%%%%%%%%%%%%%%%%%%%%%%%%%%%%%%%%%%%%%%%%%%%%%%%%%%%%%%%%%%%%%%%%%

\title{Measurement of the elliptic flow of $^3$He and $^3_\Lambda$H in Pb--Pb collisions at \mbox{$\sqrt{s_{\rm NN}} = 5.36$ TeV}}
\ShortTitle{$v_{2}$ of A=3 (hyper)nuclei in Run 3}

\Collaboration{ALICE Collaboration\thanks{See Appendix~\ref{app:collab} for the list of collaboration members}}
\ShortAuthor{ALICE Collaboration}

\begin{abstract}
The first measurement of the elliptic flow coefficient of (anti)${}^3_{\Lambda}$H and the study of the $v_2$ of $^3\overline{\mathrm{He}}$ measured in Pb--Pb collisions at \mbox{$\sqrt{s_{\rm NN}} = 5.36$ TeV} with the ALICE detector are presented. Based on the large data sample of approximately five billion events collected in 2023 during the LHC Run 3 data taking, these measurements provide important insights into the production mechanism of (anti)(hyper)nuclei, as well as into the phase-space distributions of nucleons and hyperons produced in heavy-ion collisions. The results are discussed in the context of hydrodynamic and coalescence models, highlighting how the measurement of the elliptic flow of nuclei, such as helium and hypertriton, provides critical constraints on hadronization models.
\end{abstract}
\end{titlepage}

\setcounter{page}{2}

%%%%%%%%%%%%%%%%%%%%%%%%%%%%%%%%%%%%%%%%%%%%%%%%%%%%%%%%%%%%%%%%%%%%%%%%%%%%%%%%%%%%%%%%%%%%%%%%%%%%%%%%%%%%%%%%%%%%%%%%%%%%%%%%%%%%%%%%%%%%%%%%%%%%%
\section{Introduction}
\label{sec:introduction}

In ultrarelativistic heavy-ion collisions, the formation of a strongly interacting quark–gluon plasma (QGP) is followed by its hydrodynamic expansion and subsequent hadronization into observable final-state particles. The collective dynamics of the system can be experimentally probed through anisotropic flow, which reflects the azimuthal anisotropy of the momentum distribution of produced hadrons in the transverse plane~\cite{AnisotropyAndFlow}.

In non-central collisions (i.e., impact parameter $b>0$), the initial nuclear overlap region exhibits an anisotropic geometry in the transverse plane. During the QGP expansion, the corresponding spatial eccentricities generate anisotropic pressure gradients that convert the initial coordinate-space asymmetries into measurable momentum-space anisotropies of the emitted particles. Among the various anisotropic components of the collective expansion, elliptic flow — associated with the elliptic shape of the overlap region — constitutes the dominant contribution in non-central collisions.

Anisotropic flow can be quantified through a Fourier decomposition of the azimuthal particle distribution with respect to an appropriate symmetry plane of the collision. The resulting harmonic coefficients $v_n$ characterize the collective response of the medium to the initial geometry and event-by-event fluctuations, and thus provide sensitive probes of the transport properties and the equation of state of the QGP~\cite{Voloshin:2008dg, Ollitrault:2009ie, Alver:2010gr, Qiu:2011iv}.
The present study focuses exclusively on the second-order harmonic coefficient $v_2$, which quantifies elliptic flow.

Studying the elliptic flow of different particle species provides additional information about their production mechanism as well as their interaction during the late hadronic rescattering phase. Of particular interest is the study of $v_2$ for (anti)nuclei (consisting of nucleons) and (anti)hypernuclei (composed of nucleons and hyperons). Their production mechanism is usually described by phenomenological models, such as those based on a baryon-coalescence approach~\cite{Coalescence3,Coalescence1, SATO1981153, Csernai:1986qf, Coalescence2,Blum:2017qnn,iEBE_VISHNU,Mahlein:2023fmx,Mahlein:2025bla}, according to which the formation probability of hadrons is calculated by folding the phase-space distributions of the constituent nucleons with the Wigner density of the bound state~\cite{Wigner:1932eb}.
In heavy-ion collisions, the measurement of the production of (hyper)nuclei and their subsequent flow provide unique insights into the dynamics of the system and the underlying particle production mechanism. Hydrodynamic models~\cite{Huovinen:2001cy} attribute the elliptic flow of composite nuclei primarily to their mass and transverse momentum, assuming collective expansion and thermal equilibrium of the medium. Conversely, coalescence models emphasize the role of the baryon-emitting source geometry and the spatial and momentum correlations among constituent particles, particularly for loosely bound systems like the hypertriton ($^3_{\Lambda}\mathrm{H}$). As such, comparisons between the flow behavior of light nuclei with similar masses but potentially different internal structures offer a unique test of these competing descriptions of hadronization and freeze-out dynamics~\cite{Vozabova:2024jum}.

In the context of quark-coalescence models, it is well established that the process of forming hadrons from constituent quarks inherently introduces higher-order harmonic components in the azimuthal-angle distributions at intermediate and high $p_{\mathrm T}$, as shown in Refs.~\cite{Kolb:2004gi, QuarkCoalescence}. These higher-order terms arise naturally due to the non-linear mapping from quark to hadron momentum space. The contribution of such higher harmonics becomes significant when the elliptic flow of the constituent quarks is sufficiently large.
An analogous mechanism may apply to the formation of light nuclei via nucleon coalescence. In particular, for $^3\mathrm{He}$ at high transverse momentum, where its constituent protons lie in the momentum range exhibiting maximal proton elliptic flow, the resulting composite system can inherit non-negligible higher-order modulations relative to the second-order symmetry plane $\Psi_2$. As a result, the standard Fourier decomposition, often truncated at second order, may become inadequate for describing the measured $v_2$. These higher-order harmonics, introduced by the coalescence mechanism, can affect the extracted flow observables and should be accounted for in their interpretation. Understanding their impact is therefore crucial for correctly interpreting large observed $v_2$ values with these methods and for assessing the limitations of simplified flow models.

In this work, the first measurement of the elliptic flow ($v_2$) of $^3\mathrm{He}$ and $^{3}_{\Lambda}\mathrm{H}$ in Pb--Pb collisions at $\sqrt{s_{\mathrm{NN}}} = 5.36$ TeV is reported. These results are crucial for understanding the interplay between collective flow of nucleons and the coalescence mechanism in the formation of composite particles.
The results reported in this Letter complement the extensive body of flow measurements for other hadrons and light nuclei carried out by the ALICE Collaboration~\cite{FlowChargedParticlesLHC,  Acharya:2017dmc, protonV2, Acharya:2019ttn, deuteronFlowALICE502TeV, Acharya:2020lus, flowIDhadronsXeXe, ALICE:2022zks, ALICE:2024say}.

\section{Experimental setup and data sample}
\label{sec:Experiment}
The data analyzed in this study were collected with the upgraded ALICE detector~\cite{ALICE:2023udb} during the 2023 heavy-ion run at the CERN LHC. The analysis is based on a data sample comprising approximately $5 \times 10^{9}$ minimum-bias Pb--Pb collision events.

A detailed description of the ALICE apparatus and its performance can be found in Ref.~\cite{ALICE:2023udb}. In the following, only the sub detectors relevant to this analysis are briefly presented.

Charged-particle trajectories are reconstructed in the ALICE central barrel, which covers the pseudorapidity range $|\eta| < 0.9$, using the Inner Tracking System (ITS)~\cite{ALICE:2013nwm} and the Time Projection Chamber (TPC)~\cite{ALICETPC:2020ann, CERN-LHCC-2013-020}, upgraded for the LHC Run 3 data taking.
These detectors are situated inside a solenoidal magnet that generates a uniform magnetic field of 0.5~T along the beam axis.

The ITS consists of seven cylindrical layers of silicon pixel detectors, positioned at radial distances between 2.3 and 39.3~cm from the beam axis. It is mainly used for determining primary and secondary vertices and for charged-particle tracking, but it can also be used for particle identification (PID), using the cluster size. Surrounding the ITS, the TPC is a large-volume cylindrical detector equipped with Gas Electron Multiplier (GEM) readout technology, covering the region $85 < r < 247$~cm and $-250 < z < 250$~cm in radial and longitudinal directions, respectively. The TPC enables charged-particle track reconstruction, momentum measurement, and PID through the specific energy loss (d$E$/d$x$) of particles in the gas.
The Fast Interaction Trigger (FIT)~\cite{Antonioli:1603472, Trzaska:2017reu} detector provides the interaction trigger, acts as an online and offline luminometer, offers an estimate of the vertex position and particle multiplicity at forward rapidity, and it is used to determine the event plane. FIT consists of different subsystems, i.e., the FT0, made out of two arrays of fast Cherenkov radiators placed symmetrically around the interaction point, and the FV0 and the FDD subsystems, which are scintillator arrays. These detectors are used for the measurement of particle multiplicity at forward (backward) rapidity. Specifically, FT0 comprises two modules (FT0A and FT0C) located along the beam pipe: the FT0A at 330~cm from the interaction point ($3.5 < \eta < 4.9$) and the FT0C at -84.3~cm ($-2.1 < \eta < -3.3$)~\cite{ALICE:2023udb}. The FT0 is used as the reference detector for the event-centrality determination in the present analysis.

\subsection{Event and candidate selection}
Collision events are required to have a reconstructed primary vertex located within $\pm 10$~cm of the nominal interaction point along the beam axis. Events having incomplete detector information and events with pile-up were excluded.
These selection steps result in a high-quality data sample of about 5 billion collision events, well-suited for flow measurements of rare objects such as (hyper)nuclei. This study examines different centrality intervals: for the hypertriton elliptic flow, 0--20\% and 20--60\%, whereas for that of $^3\mathrm{He}$, 0--10\%, 10--20\%, 20--30\%, 30--40\%, and 40--60\%.

The analysis of $^3\mathrm{He}$ is carried out focusing on antimatter nuclei only ($^3\overline{\mathrm{He}}$), while for $^{3}_{\Lambda}\mathrm{H}$ matter and antimatter nuclei are summed up. This is possible since previous analyses from Run 2 demonstrated that both the production yields and the anisotropic flow are the same for matter and antimatter~\cite{FlowChargedParticlesLHC, protonV2, Acharya:2017dmc}. Unlike its matter counterpart, the $^3\overline{\mathrm{He}}$ originates directly from primary interactions, without any background originating from knock-out interactions in the detector. For hypertriton, instead, the contribution from knockout is negligible due to its strangeness content, thus both matter and antimatter are analysed to obtain a larger dataset of candidates. Any residual background is removed using a fitting procedure, as described below. Hereinafter, $^{3}_{\Lambda}\mathrm{H}$ will refer to the sum of matter and antimatter.

The $^3\overline{\mathrm{He}}$ candidates are identified through the measurement of their specific energy loss (d$E$/d$x$) in the TPC, expressed in terms of $n\sigma_{\mathrm{TPC}}$, which is defined as the difference between the measured d$E$/d$x$ and the expected value for that particle species at a given momentum, divided by the TPC resolution (in terms of standard deviations). Since $^3\overline{\mathrm{He}}$ is characterized by a charge number $|Z|=2$, it can be well separated from other particle species, in a broad momentum range (from 2 to 8 GeV/$c$).
The small contamination from other particles with $|Z|=1$ at low transverse momentum ($p_{\mathrm{T}}$) can be drastically suppressed by requiring a selection on the average ITS cluster size. Indeed, the average size of clusters of firing pixels can be used as a proxy of specific energy loss, hence to identify heavily ionizing particles such as (hyper)nuclei, as originally proposed in Ref.~\cite{ALICE:2013nwm}. In particular, nuclei with $|Z|=2$, at a given momentum, lose more energy than $|Z|=1$ particles and consequently create a cluster of pixels with a larger average size.
Hence, in order to use the pixel cluster-size response of the ITS as a PID observable, the $n\sigma_{\mathrm{ITS}}$ is built, similarly to the case of the PID using the energy loss in the TPC.
For each track, the cluster-size estimator is given by the truncated mean of the cluster sizes measured in the seven ITS layers. For a given particle hypothesis (in the case of this analysis, the $^3\overline{\mathrm{He}}$), this measured estimator is compared to the expected cluster-size response at the same momentum and geometry (pseudorapidity and incidence angle), and the difference is expressed in units of the corresponding detector resolution. The expected response and resolution are calibrated in narrow momentum bins using Monte Carlo simulations by fitting the cluster-size distributions and then smoothing the extracted parameters with respect to kinematics.
Candidates are selected by requiring $n\sigma_{\mathrm{ITS}} > -1$ and $|n\sigma_{\mathrm{TPC}}| < 2$. Additionally, tracks are required to have at least 110 clusters in the TPC, to provide a good PID. With these selections, the purity of the sample of $^3\overline{\mathrm{He}}$, estimated via a fit to the $n\sigma_{\mathrm{TPC}}$, is above 99\%. The selected sample contains also a contribution coming from the weak decay of hypertriton, which accounts for less than 1\% of the total yields and is negligible.

Hypertriton candidates are reconstructed via their two-body mesonic decay channel~\cite{ALICE:2022sco}:
\[
^{3}_{\Lambda}\mathrm{H} \rightarrow ^{3}\mathrm{He} + \pi^- \, \mathrm{and} \; \mathrm{c.c.}
\]
The weakly-decaying hypernuclei are reconstructed with a \mbox{V$^0$-like} two-body secondary-vertex finder, as used in the previous ALICE hypertriton analysis~\cite{hypertriton_pPb_ALICE}. The algorithm utilizes the opposite-sign decay daughters to reconstruct the mother particle, fits a common displaced decay vertex, and applies topological selections based on the daughter distances of closest approach (DCA), the decay length, and the pointing angle of the reconstructed mother candidate to the primary vertex. The new version of the secondary-vertex finder has been adapted to cope with the continuous readout: first, the tracks reconstructed with ITS and TPC are paired only if they fall in a time window compatible with the same collision, then the decay vertices are reconstructed and finally validated by checking the pointing to the primary vertex~\cite{Rohr:2019ava}.

The daughter tracks are required to have a pseudorapidity within $|\eta|<0.9$ and the following transverse momentum selections: $p_{\rm T}^{^{3}\rm{He}} > 2$~GeV/$c$ and $p_{\rm T}^{\pi} < 1.5$~GeV/$c$.
The former selection is applied to reduce the contamination from the knock-out interactions; the latter is applied to avoid contamination from kaons and protons, which becomes considerable above 1.5 GeV/$c$.
The daughter tracks are identified using TPC and ITS, as done for the $^3\overline{\mathrm{He}}$. In particular, $|n\sigma_{\mathrm{TPC}}| < 4$ is required, in a track-by-track approach for the $\pi$ and the $^3$He daughters, in addition to a selection of the average cluster size of the $^3$He in the ITS greater than 5. Requiring the daughter $^{3}\mathrm{He}$ to satisfy the ITS cluster-size selection strongly suppresses the combinatorial background and constitutes the primary handle for hypertriton identification. Additionally, $^3$He candidate daughter tracks are required to have more than 110 clusters in the TPC.
Finally, a rather stringent cut on the cosine of the angle between the total momentum of the daughter tracks at the secondary vertex and the vector connecting the primary and secondary vertex (pointing angle) was applied, requiring that cos($\theta_{pointing}) >$ 0.99985. The proper decay length, expressed as $ct$, was required to exceed 1~cm.

\section{Data analysis}
\label{sec:DataAnalysis}
The hypertriton signal is extracted from the invariant-mass distribution of the decay products. The invariant mass is calculated from the measured momenta of the daughter tracks. The distribution is modeled using a Double-Sided Crystal Ball (DSCB) function for the signal, with the tail parameters constrained by Monte Carlo simulations, and a second-order polynomial to describe the combinatorial background. Conversely, for the $^3\overline{\mathrm{He}}$ there is no explicit signal extraction as the selected candidates in the $n_{\sigma_{\mathrm{TPC}}}$ distributions are extremely pure, as discussed in the previous Section.

\begin{figure}[htb]
    \centering
    \includegraphics[width=0.49\textwidth]{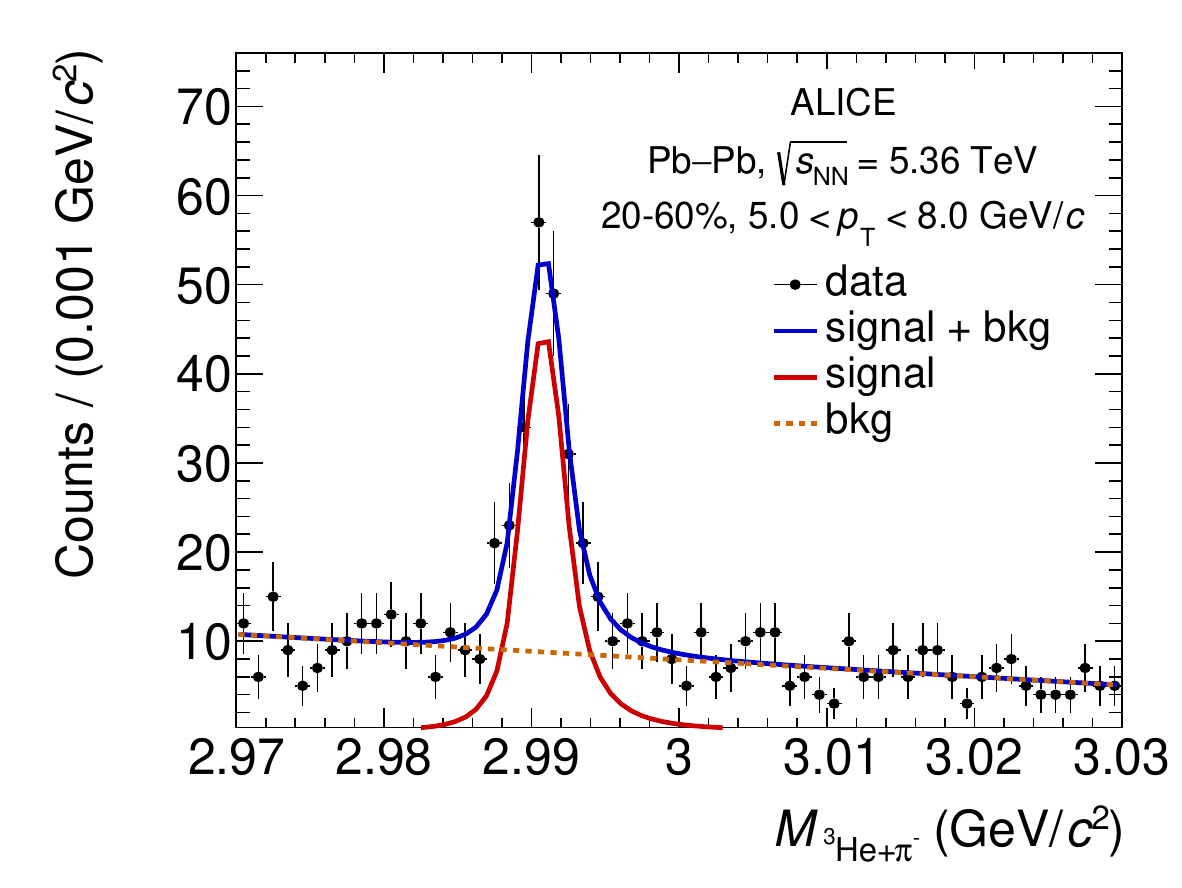}
    \includegraphics[width=0.49\textwidth]{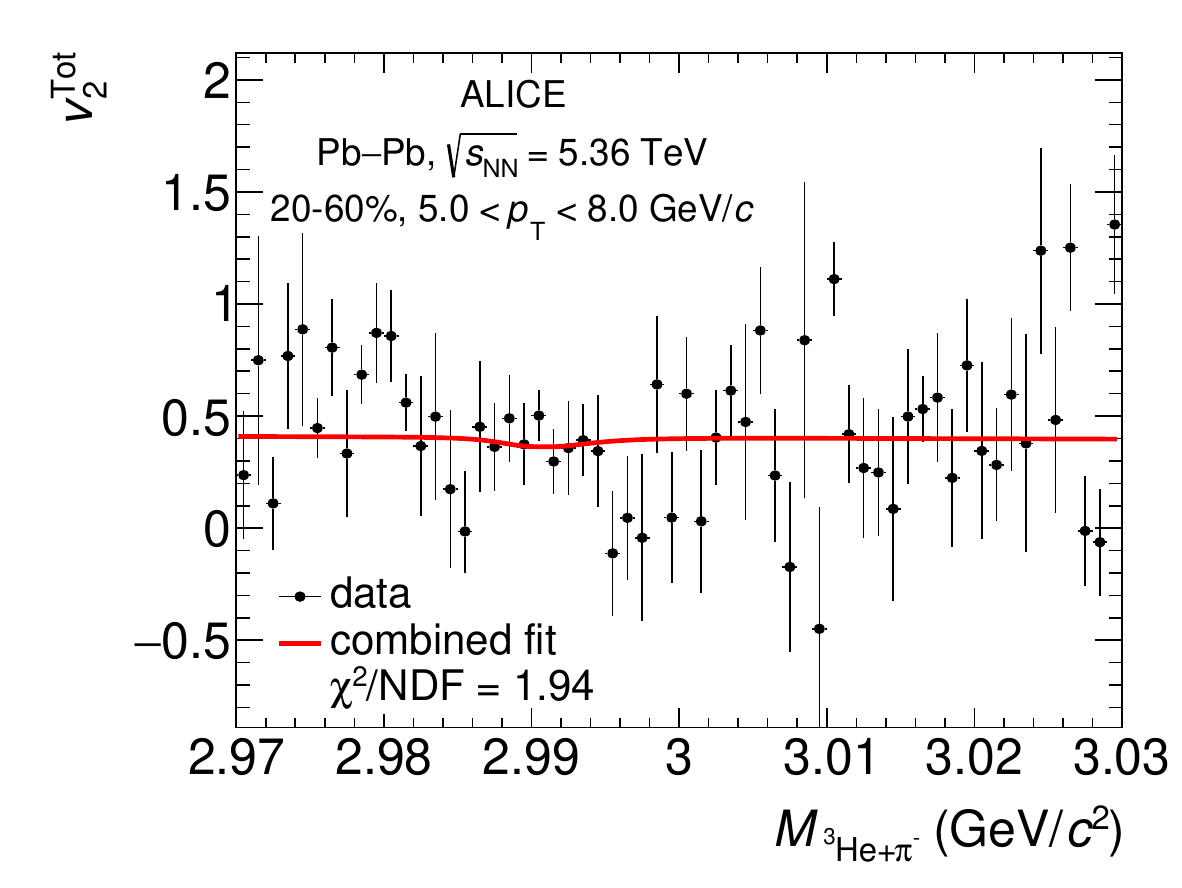}
    \caption{Invariant-mass distribution of $^{3}_{\Lambda}\mathrm{H}$ candidates (left) and $v_{2}$ as a function of the invariant mass (right), in the $5<p_{\rm T}<8$ GeV/$c$ transverse-momentum interval and the 20--60\% centrality interval. In the left panel, the solid red line represents the fit with a DSCB function (signal) while the dotted line displays the fit of the background with a second-order polynomial function. The total fit is shown with the solid blue line. In the right panel, the solid red line represents the fit function described in Eq.~\ref{eq:v2tot}.}
    \label{fig:invariant_mass}
\end{figure}

The left panel of Fig.~\ref{fig:invariant_mass} displays the invariant-mass distribution of the hypertriton candidates for the 20--60\% centrality interval in the $5<p_{\rm T}<8$ GeV/$c$ transverse-momentum interval, with the fit functions describing the signal, the background, and the total distribution with different colors.

The reconstruction efficiency of hypertritons, defined as the ratio between the reconstructed and generated candidates, depends on $p_{\mathrm T}$ and centrality, with a mild decrease at low $p_{\mathrm T}$ and a stronger centrality dependence (from about 5\% in the 0–10\% interval to about 15\% in the 50–60\% one). To account for these variations, a centrality-dependent reweighting is implemented using $p_{\mathrm T}$-dependent weights from a Blast-Wave function~\cite{BlastWave1}. A Blast-Wave parametrisation of the hypertriton transverse-momentum spectra in Pb--Pb collisions at $\sqrt{s_{\mathrm{NN}}}=5.02~\text{TeV}$~\cite{ALICE:2024koa} is used as weight to reflect the expected $p_{\mathrm T}$ distribution of the signal when averaging the efficiency. Then, such a centrality-dependent weight is applied candidate-by-candidate, so that the Blast-Wave weights affect the efficiency correction rather than the measured $v_{2}$.

The azimuthal distribution of (hyper)nuclei produced in the collision with respect to the \textit{n}-th order flow symmetry plane $\Psi_{n}$~\cite{Voloshin:2008dg,Ollitrault:2009ie,Alver:2010gr,Qiu:2011iv} can be expressed as a Fourier series
 \begin{equation}
 E \frac{\mathrm{ d^3} N}{\mathrm{d} p^3} = \frac{1}{2\pi} \frac{\mathrm {d^2} N}{p_\mathrm{T} \mathrm{d} p_{\rm{T}} \mathrm{d} y } \left( 1 + \sum_{n=1}^{\infty} 2 v_n \cos \left( n \left( \varphi -  \Psi_{n} \right) \right) \right),
\label{eq:flowyield}
\end{equation}
where $E$ is the energy of the particle, $p$ its momentum, $\varphi$ the azimuthal angle, $y$ the rapidity, and
\begin{equation}
v_n = \langle \cos \left( n(\varphi - \Psi_{n}) \right)\rangle \, .
\end{equation}
The $v_{2}$ coefficient is measured using the Scalar Product (SP) method~\cite{Adler:2002pu,Voloshin:2008dg}, as done in Refs.~\cite{Acharya:2017dmc,deuteronFlowALICE502TeV, ALICE:2024say}. The $Q$-vector is computed from the azimuthal distribution of a set of charged particles (so called reference flow particles) as
\begin{equation}
    \textbf{\textit{Q}}_n = \sum w_i e^{in\varphi_i} ,
\end{equation}
where $w_i$ is a weight applied to correct for non-uniform acceptance and efficiency of the detector (as done in Refs.~\cite{FlowChargedParticlesLHC, Acharya:2020lus, protonV2}), $n$ is the order of the harmonic, and $\varphi_i$ is the azimuthal angle for the $i$th reference flow particle.

For the generic $n$ harmonic number, the Fourier coefficient is obtained as:

\begin{equation}
    v_n \{\mathrm{SP}\} = \frac{\left\langle \langle \textbf{u}_{n,k}{\textbf{Q}^*_n} \rangle \right\rangle}{\sqrt{\frac{\langle \textbf{Q}_n \textbf{Q}^{A*}_n \rangle \langle \textbf{Q}_n \textbf{Q}^{B*}_n \rangle} {\langle \textbf{Q}^A_n \textbf{Q}^{B*}_n \rangle}}}
\label{eq:flow}
\end{equation}
where \textbf{u}$_{n,k}$ = $e^{(in\varphi_k)}$ is the unit flow vector of the particle of interest $k$ with azimuthal angle $\varphi_k$, and $\textbf{Q}_n$ is the flow vector.
The double brackets $\langle \langle ... \rangle \rangle$ denote an average over all particles in all events, and $*$ the complex conjugate.
The denominator of Eq.~\ref{eq:flow} is a normalization factor that accounts for the magnitude of the reference flow. In this analysis, the vector \textbf{Q}$_{n}$ is calculated from the azimuthal distribution of the energy deposit in the FT0C. The $\textbf{Q}_{n}^A$ and $\textbf{Q}_{n}^B$ are determined from the azimuthal distribution of the energy deposit in the FT0A and the azimuthal distribution of the tracks reconstructed in the TPC, respectively. Using this configuration, a pseudorapidity gap of $|\Delta\eta|>1.3$ between the (hyper)nucleus of interest and the reference flow particles is introduced. Such a gap reduces the non-flow effects due to correlations not arising from the collective expansion of the system, e.g. resonance decays and jets~\cite{ALICE:2018rtz}.

Given the extremely pure sample of ${}^{3}\mathrm{He}$, its $v_2$ is directly taken as the mean value of the $v_2$ distribution for a given $p_{\mathrm{T}}$ interval. On the contrary, the contribution to the measured elliptic flow ($v_2^{\mathrm{Tot}}$) from the $v_2$ of random ${}^{3}\mathrm{He}-\pi$ pairs with a mass consistent with the hypertriton mass ($v_2^{\mathrm{Bkg}}$) has to be removed. This can be done by studying the measured $v_2^{\mathrm{Tot}}$ as a function of the invariant mass and, leveraging the fact that the $v_2^{\mathrm{Tot}}$ is additive, one can factorize the contribution of signal ($v_2^{\mathrm{Sig}}$) and background ($v_2^{\mathrm{Bkg}}$) as follows:

\begin{equation}
v_2^{\mathrm{Tot}} (m) = v_2^{\mathrm{Sig}}(m) \frac{N^{\mathrm{Sig}}}{N^{\mathrm{Tot}}}(m) + v_2^{\mathrm{Bkg}} (m) \frac{N^{\mathrm{Bkg}}}{N^{\mathrm{Tot}}}(m),
\label{eq:v2tot}
\end{equation}
where $N^{\mathrm{Tot}}$ is the total number of candidates, $N^\mathrm{Bkg}$ and $N^\mathrm{Sig}$ = $N^\mathrm{Tot}$ - $N^\mathrm{Bkg}$ are the number of background and signal counts for a given mass and \pt interval.
The invariant-mass dependent elliptic flow, $v^{\mathrm{Tot}}_2(m)$, is obtained in each $p_{\mathrm{T}}$ and centrality interval and fitted simultaneously with the corresponding invariant-mass spectrum. The signal and background yields $N_{\mathrm{Sig}}(m)$ and $N_{\mathrm{Bkg}}(m)$ in Eq.~\ref{eq:v2tot} are taken from the fit of the invariant-mass distribution with the functions mentioned above, which define the mass-dependent signal and background fractions. In the fit to $v^{\mathrm{Tot}}_2(m)$, the hypertriton elliptic flow $v^{\mathrm{Sig}}_2$ is assumed to be independent of the invariant mass, while the background elliptic flow $v^{\mathrm{Bkg}}_2(m)$ is parameterised as a first-order polynomial in $m$. A common $\chi^2$ minimisation of the mass and $v_2(m)$ distributions is performed, yielding $v^{\mathrm{Sig}}_2$ in each $p_{\mathrm{T}}$ interval. An example of the simultaneous fitting procedure is shown in the right panel of Fig.~\ref{fig:invariant_mass}.

\subsection{Systematic Uncertainties}
The systematic uncertainties arise from the track selection criteria, the choice of the background fit function, and the procedure employed to extract the flow.

To estimate the uncertainty associated with track selection criteria, a systematic variation is performed by simultaneously changing all relevant selection variables, such as PID and track quality parameters, within intervals of $\pm 50\%$, and, in the case of hypertriton, also minimum $p_{\rm T}$ within (2.0,2.5) GeV/$c$, and the cosine of the pointing angle within (0.9998,0.9999). The variation range of $\pm 50\%$ around the nominal values is chosen to span a broad set of selection choices, covering potential dependencies of the result on the specific cut configuration, while avoiding excessively loose cuts that would introduce a large background contribution. For each of the 200 (50 for hypertriton) randomized trials, the full analysis chain is repeated, and the relative uncertainty is calculated as the RMS of the $v_2$ distribution in each \pt bin. The Barlow test has been performed, and the variations below 2$\sigma$ have been rejected~\cite{Barlow:2002yb}. For the $^3\overline{\mathrm{He}}$ analysis, the systematic uncertainties are evaluated separately for each centrality interval, and their contribution is always below 1\%. On the contrary, for hypertriton the systematic uncertainties are evaluated using the full 0–60\% centrality interval to minimize statistical effects. The contribution related to track selections is of about $10\%$, only mildly depending on the \pt.

For the hypertriton analysis, another significant source of uncertainty stems from the fit function used to model the background in the invariant-mass distributions. To evaluate this, three functional forms, i.e., a second-order polynomial (default), a first-order polynomial, and an exponential function, are tested. The systematic uncertainty is taken as half the maximum variation among the $v_2$ values obtained with these fits. For the lowest \pt bin, the result from the first-order polynomial is excluded due to its poor background description. This procedure is applied separately for each centrality interval (0–20\% and 20–60\%), and the resulting uncertainties range between $8\%$ at low \pt and $1\%$ at high \pt in the 0-20\% centrality interval, and they are about $1\%$ in the 20-60\% interval.

Moreover, in order to exclude the presence of short-range non-flow effects, the analysis has been repeated varying the pseudorapidity gap between the measurement and the reference regions, passing from $|\Delta\eta| > 1.3$ to $|\Delta\eta| > 2.1$. The two analyses yield compatible results, excluding non-flow effects.

Finally, a contribution related to the estimation of the Q-vectors is taken into account. Both analyses have been carried out using two different methods for the determination of the event Q-vectors, which differ in the corrections applied to the raw values: in one case twisting~\cite{Selyuzhenkov:2007zi} is applied, in the other it is not. Such uncertainty is estimated for the $^3\overline{\mathrm{He}}$ analysis, and then applied also to the hypertriton after interpolating the values to match the larger \pt intervals of the hypertriton. The final contribution from this method is approximately 5\%, with a mild dependence on \pt and centrality.

The final systematic uncertainty is obtained by summing the individual contributions from each of these sources in quadrature, under the assumption that they are uncorrelated.

\section{Results}
\label{sec:Results}

\begin{figure}[htb]
    \centering
    \includegraphics[width=0.45\textwidth]{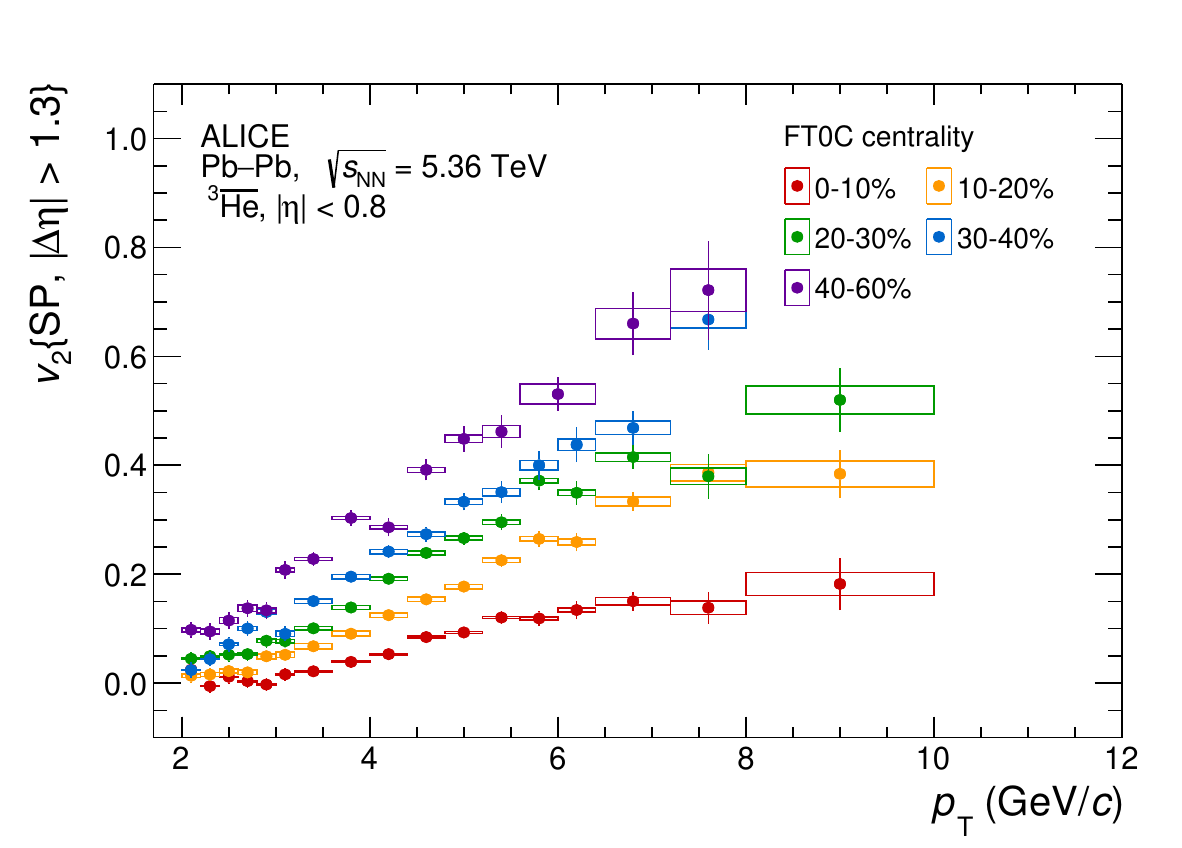}
    \includegraphics[width=0.45\textwidth]{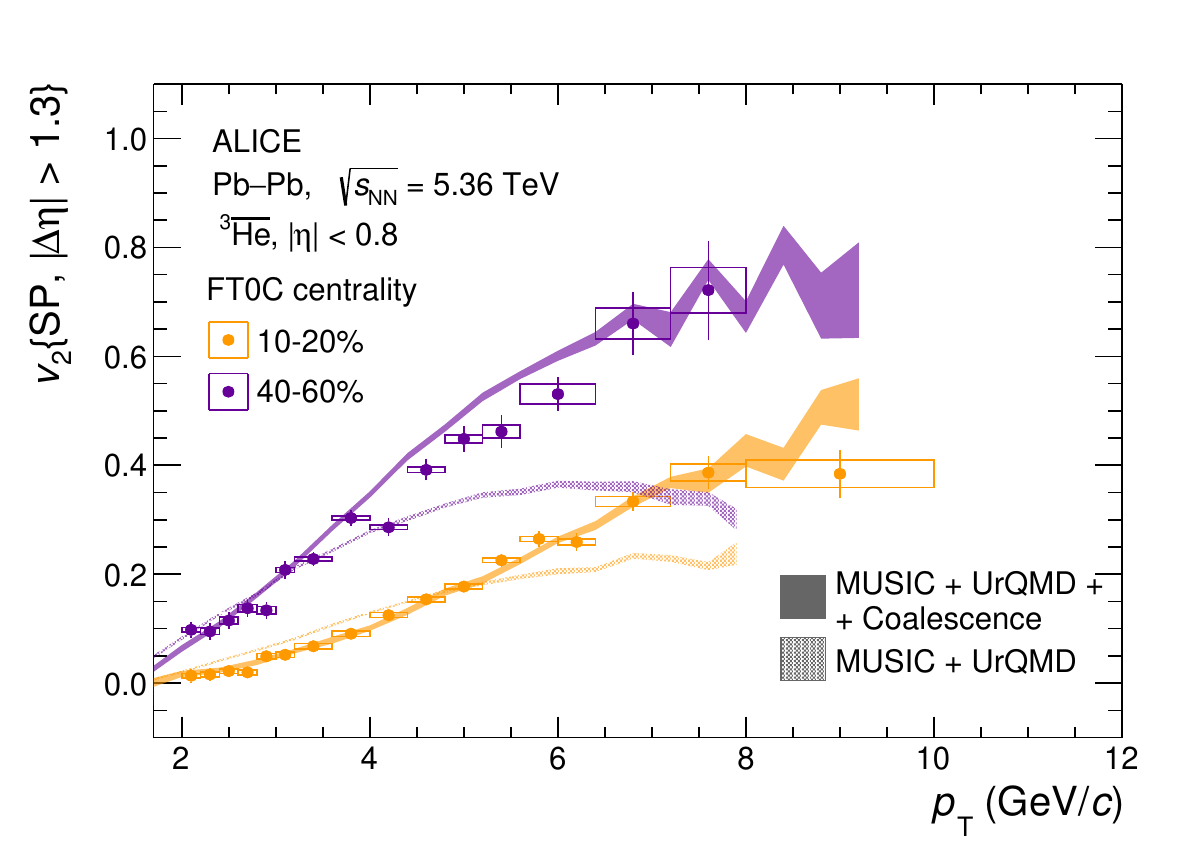}
    \caption{Measured $v_2$ of $^3\overline{\mathrm{He}}$ as a function of $p_{\mathrm T}$ for the various centrality intervals (left) and comparison with predictions of hydrodynamical models~\cite{Ma:2026wtv} with and without a coalescence afterburner for two selected centrality intervals (right). Error bars represent statistical uncertainties, while boxes indicate systematic uncertainties, respectively. The widths of the model bands represent their statistical uncertainties.}
    \label{fig:v2_results}
\end{figure}

Figure~\ref{fig:v2_results} shows the measured $v_2$ of $^3\overline{\mathrm{He}}$ (left panel) as a function of transverse momentum for the different centrality intervals.
The $v_2$ values exhibit an increase with $p_{\mathrm T}$, which at low--intermediate momenta is commonly interpreted as a consequence of the medium collective anisotropic expansion converting the initial spatial eccentricity into a momentum-space anisotropy. In addition, the magnitude of $v_2$ increases from central to more peripheral collisions, consistent with the larger initial geometric eccentricity of the overlap region in peripheral events.

In the right panel of Fig.~\ref{fig:v2_results}, the data in two centrality intervals (10--20\% and 40--60\%) are compared with the expectations of hydrodynamical calculations performed within two otherwise identical multi-stage hybrid frameworks~\cite{Ma:2026wtv}, which differ only in the treatment of the formation of light nuclei. Up to hadronization, in both approaches the space--time evolution of the quark--gluon plasma is described by the (3+1)-dimensional relativistic viscous hydrodynamics model MUSIC~\cite{Paquet:2015lta,Shen:2017bsr,Shen:2020mgh}, initialized with 3D collision-geometry-based initial conditions from the IP-Glasma framework~\cite{Shen:2020jwv}. A crossover equation of state at finite baryon density (NEOS-BQS)~\cite{Monnai:2019hkn} is employed. Particle production proceeds via the Cooper--Frye prescription~\cite{Cooper:1974mv} on a constant energy-density hypersurface, followed by hadronic rescattering and decays modeled with UrQMD~\cite{Bass:1998ca}.

The two calculations differ only in the treatment of nuclei after hadronization. In the prompt-production scenario, $^3\overline{\mathrm{He}}$ is produced directly at hadronization, together with all other hadrons, and is subsequently propagated through the hadronic phase. In the alternative scenario, a coalescence afterburner is applied at kinetic freeze-out, following the model of Ref.~\cite{Liu:2024ygk}, which employs a Wigner-function-based coalescence approach as implemented in Ref.~\cite{iEBE_VISHNU}. In this framework, $^3\overline{\mathrm{He}}$ and hypertriton nuclei are formed from the emitted nucleons. The $^3\overline{\mathrm{He}}$ wave function is parametrized as a harmonic-oscillator state with radius $R = 1.76$~fm and statistical factor $g = 1/4$, which accounts for the spin combinations of the constituent nucleons.

An interesting feature observed in the measured $v_2$ of $^3\overline{\mathrm{He}}$ at high transverse momentum ($p_{\mathrm{T}} > 6$~GeV/$c$) in peripheral collisions (see Fig.~\ref{fig:v2_results}) is that $v_2$ exceeds 0.5. According to Eq. ~\ref{eq:flowyield}, truncating the Fourier expansion at second order implies that such high $v_2$ values would yield negative particle counts in certain azimuthal regions, which is unphysical. This suggests that higher-order harmonics with respect to the second-order symmetry plane may become non-negligible when measuring elliptic flow. In fact, the presence of higher harmonics like $v_{4}\{\Psi_{2}\}$ naturally arises in hydrodynamic evolution via non-linear mode coupling (typically scaling as $(v_2)^2$), but their magnitude is usually small~\cite{Yan:2015jma}.
As shown in the right panel of Fig.~\ref{fig:v2_results}, only the calculation including the coalescence afterburner is able to reproduce the magnitude of the measured $v_2$, while the prompt-production scenario does not describe the observed trend. A difference between the two predictions is expected, since the coalescence process introduces a non-linear mapping from nucleons to nuclei, which can modify the azimuthal distributions with respect to those of the constituent nucleons. Similar effects have been observed in quark-coalescence models, where the recombination of partons into hadrons generates higher-order harmonic contributions in the azimuthal distributions at intermediate and high transverse momentum~\cite{Kolb:2004gi, QuarkCoalescence}. By analogy, in nucleon coalescence, the formation of $^3\overline{\mathrm{He}}$ at high $p_{\mathrm{T}}$, where proton elliptic flow is larger, can similarly induce additional higher-order modulations in the azimuthal structure of the composite nucleus.
Possible contributions from long-range back-to-back jet correlations are expected to be strongly suppressed for light (hyper)nuclei, whose production is dominated by soft bulk processes and coalescence.

It should also be noted that the measured azimuthal modulation may receive contributions not only from the second-order flow component associated with the $\Psi_2$ symmetry plane, but also from genuine higher-order flow effects, including contributions correlated with $\Psi_4$. Disentangling these effects from those induced by the coalescence mechanism is beyond the scope of this work and will be the subject of future studies.

\begin{figure}[htb]
    \centering
    \includegraphics[width=0.49\textwidth]{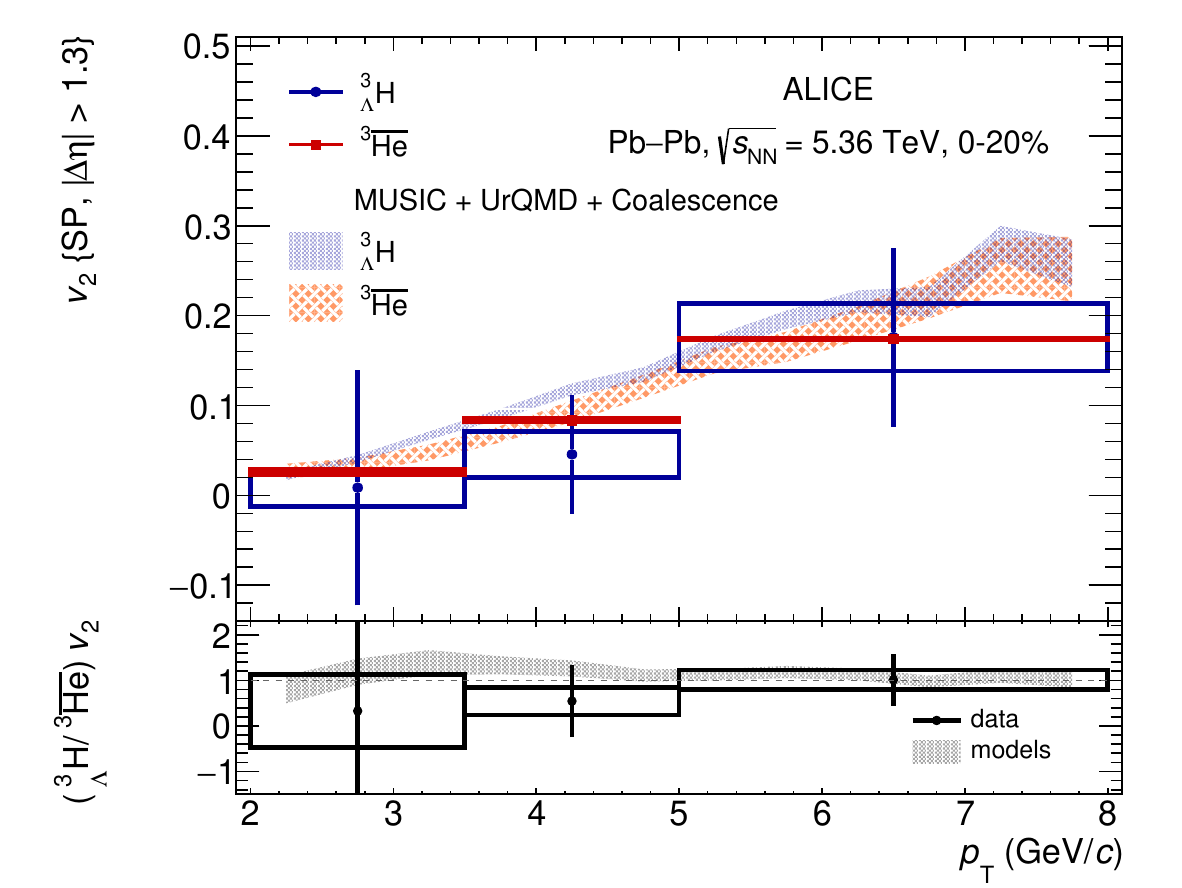}
    \includegraphics[width=0.49\textwidth]{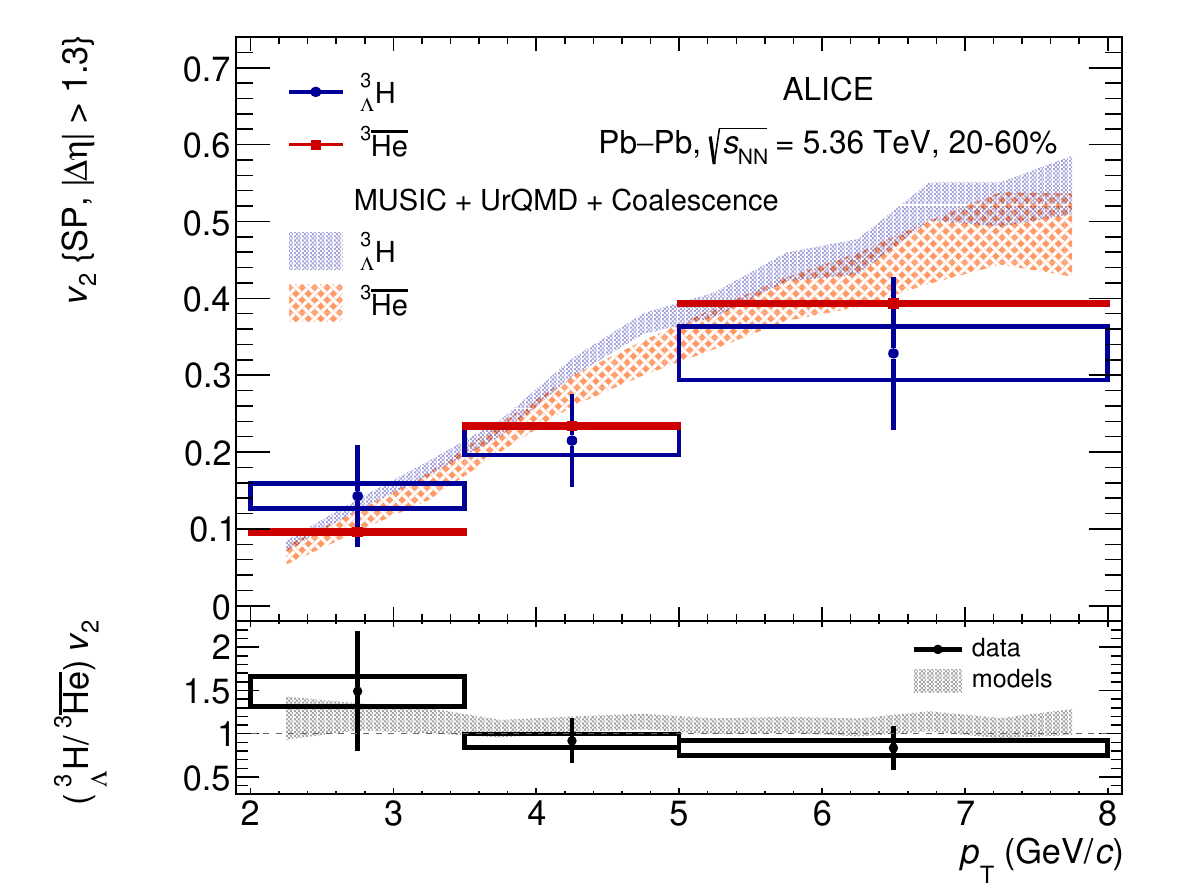}
    \caption{$v_2$ of $^3\overline{\mathrm{He}}$ (red markers) and hypertriton (blue markers) as a function of $p_{\mathrm T}$ for the 0-20\% (left panel) and 20-60\% (right panel) centrality intervals. Error bars represent statistical uncertainties, while boxes indicate systematic uncertainties, respectively. The data points are compared with the coalescence model expectations (see the text for details) for the two nuclear species. }
    \label{fig:model_comparison}
\end{figure}

In Fig.~\ref{fig:model_comparison}, the results of the $v_2$ of the $^3_{\Lambda}\mathrm{H}$ are shown, and compared with those of $^3\overline{\mathrm{He}}$, using the same binning in transverse momentum and centrality for both, to allow for a direct comparison between the two species.
The data are compared to the predictions of the model with coalescence described above. The version without coalescence is not reported, due to the rarity of the production of hypertriton at the hadronization stage. Similarly to what is done for $^3\overline{\mathrm{He}}$, in such a model, hypertriton is formed from the coalescence of a proton, a neutron, and a $\Lambda$ hyperon. The hypertriton wave function is modeled as a double Gaussian (one Gaussian for the deuteron and one for the d-$\Lambda$ molecule), characterized by spatial parameters $\sigma_1 = 2.26$~fm and $\sigma_2 = 6.52$~fm, reflecting the extended size and weak binding of the hypertriton~\cite{Liu:2024ygk}. Finally, a statistical factor $g = 1/4$ is used. In Fig.~\ref{fig:model_comparison}, the comparison between data and model shows good a agreement, as also quantified by the ratio around the unity of the measured $v_2$ and the model predictions for the two species, shown in the bottom panels.
The same hydrodynamical model successfully reproduces the measured $v_2$ of light-flavored particles (namely, pions, kaons, protons) and composite nuclei (produced by coalescence using an afterburner), with predictions found to be in good agreement with results from the LHC Run~2.

The comparison between the elliptic flow of $^3\overline{\mathrm{He}}$ and $^3_\Lambda$H provides valuable insight into the particle production mechanisms and the properties of the emitting source in relativistic heavy-ion collisions. Although these two nuclei have comparable masses, i.e., $m(^3$He$) \approx 2.809$ GeV/$c^2$ and $m(^3_\Lambda$H$) \approx 2.991$ GeV/$c^2$, they differ significantly in their internal structure and spatial extent. The charge radius of $^3\overline{\mathrm{He}}$ is approximately 1.96~fm, while the hypertriton has a charge radius of about 5 fm~\cite{Nemura:1999qp}, driven by its extremely weak binding energy and loose spatial configuration, with a distance of the $\Lambda$ relative to the deuteron core of the order of 10~fm.

In the context of hydrodynamic models, which assume local thermal equilibrium and collective expansion of the system, the $v_2$ of composite particles is largely governed by their mass and transverse momentum. Given the similar masses of $^3\overline{\mathrm{He}}$ and $^3_\Lambda$H, these models predict similar flow behavior for the two nuclei, as the flow is primarily driven by the collective motion of the bulk medium, and the size of the particle plays a subleading role~\cite{iEBE_VISHNU,Oliinychenko:2018ugs}.
Conversely, models based on the coalescence mechanism can lead to a different interpretation. Indeed, coalescence is based on the interplay between the size of the particle-emitting source and the size of the nuclear wave function~\cite{Coalescence3,Vozabova:2024jum}.
Since hypertriton has an extremely small binding energy and a large spatial extent, its formation via coalescence depends strongly on the local phase-space density and correlations of its constituent particles at kinetic freeze-out. As a result, the measured $v_2$ may reflect the anisotropic structure of the emission source more directly than the intrinsic size of the nucleus itself.
In this sense, the elliptic flow of loosely bound states like the hypertriton may be influenced by the anisotropy of the source in- and out-of-plane~\cite{Vozabova:2024jum}, rather than by its size.

Within the current experimental uncertainties, the measured $v_2$ values of $^3\overline{\mathrm{He}}$ and $^{3}_{\Lambda}\mathrm{H}$ are compatible with each other, despite the very different internal structures and binding energies of the two nuclei. A similar behavior is observed in the coalescence-based model calculations, which predict comparable elliptic flow values for the two species. On the contrary, previous measurements~\cite{ALICE:2024koa} showed that the coalescence model predicts a difference between the yields of hypertriton and $^3$He nuclei of a factor of about 5.
This suggests that the internal structure of the nucleus plays a subleading role in determining the observed flow pattern, in contrast to what was observed for the yields. Instead, the elliptic flow appears to be driven predominantly by the collective motion of the system and by the similar total masses of the two composite particles.

\section{Summary}
\label{sec:Summary}
In this letter, the first measurement of the elliptic flow of hypertriton and the most precise measurement of the $^3\overline{\mathrm{He}}$ elliptic flow in Pb--Pb collisions at $\sqrt{s_{\mathrm{NN}}} = 5.36$ TeV are presented. The analysis is based on the large LHC Run 3 dataset collected in 2023 by the ALICE experiment. The results of this work reveal that the elliptic flow of $^3\overline{\mathrm{He}}$ and $^{3}_{\Lambda}\mathrm{H}$ exhibit similar and significant collective behaviors. The measured $v_2$ is consistent with hydrodynamic models paired with coalescence-based formation mechanisms.
The measurement of the elliptic flow of hypertriton paves the way for the measurement of its local polarization with respect to the beam axis~\cite{Sun:2025oib}, which allows the determination of its spin, a crucial input for production models but so far unmeasured.

%%%%%%%%%%%%%%%%%%%%%%%%%%%%%%%%%%%%%%%%%%%%%%%%%%%%%%%%%%%%%%%%%%%%%%%%%%%%%%%%%%%%%%%%%%%%%%%%%%%%%%%%%%%%%%%%%%%%%%%%%%%%%%%%%%%%%%%%%%%%%%%%%%%%%

%%%%% acknowledgements - handled by EB chairs
\newenvironment{acknowledgement}{\relax}{\relax}
\begin{acknowledgement}
\section*{Acknowledgements}
% add specific acknowledgements here
% ...but please don't remove the line below: funding agencies
% will be acknowledged with a custom tex file handled by EB chairs after Collaboration Round 2
% Version: 2026-03-04

The ALICE Collaboration would like to thank all its engineers and technicians for their invaluable contributions to the construction of the experiment and the CERN accelerator teams for the outstanding performance of the LHC complex.
The ALICE Collaboration gratefully acknowledges the resources and support provided by all Grid centres and the Worldwide LHC Computing Grid (WLCG) collaboration.
The ALICE Collaboration acknowledges the following funding agencies for their support in building and running the ALICE detector:
A. I. Alikhanyan National Science Laboratory (Yerevan Physics Institute) Foundation (ANSL), State Committee of Science and World Federation of Scientists (WFS), Armenia;
Austrian Academy of Sciences, Austrian Science Fund (FWF): [M 2467-N36] and Nationalstiftung f\"{u}r Forschung, Technologie und Entwicklung, Austria;
Ministry of Communications and High Technologies, National Nuclear Research Center, Azerbaijan;
Rede Nacional de F\'{\i}sica de Altas Energias (Renafae), Financiadora de Estudos e Projetos (Finep), Funda\c{c}\~{a}o de Amparo \`{a} Pesquisa do Estado de S\~{a}o Paulo (FAPESP) and The Sao Paulo Research Foundation  (FAPESP), Brazil;
Bulgarian Ministry of Education and Science, within the National Roadmap for Research Infrastructures 2020-2027 (object CERN), Bulgaria;
Ministry of Education of China (MOEC) , Ministry of Science \& Technology of China (MSTC) and National Natural Science Foundation of China (NSFC), China;
Ministry of Science and Education and Croatian Science Foundation, Croatia;
Centro de Aplicaciones Tecnol\'{o}gicas y Desarrollo Nuclear (CEADEN), Cubaenerg\'{\i}a, Cuba;
Ministry of Education, Youth and Sports of the Czech Republic, Czech Republic;
The Danish Council for Independent Research | Natural Sciences, the VILLUM FONDEN and Danish National Research Foundation (DNRF), Denmark;
Helsinki Institute of Physics (HIP), Finland;
Commissariat \`{a} l'Energie Atomique (CEA) and Institut National de Physique Nucl\'{e}aire et de Physique des Particules (IN2P3) and Centre National de la Recherche Scientifique (CNRS), France;
Bundesministerium f\"{u}r Forschung, Technologie und Raumfahrt (BMFTR) and GSI Helmholtzzentrum f\"{u}r Schwerionenforschung GmbH, Germany;
National Research, Development and Innovation Office, Hungary;
Department of Atomic Energy Government of India (DAE), Department of Science and Technology, Government of India (DST), University Grants Commission, Government of India (UGC) and Council of Scientific and Industrial Research (CSIR), India;
National Research and Innovation Agency - BRIN, Indonesia;
Istituto Nazionale di Fisica Nucleare (INFN), Italy;
Japanese Ministry of Education, Culture, Sports, Science and Technology (MEXT) and Japan Society for the Promotion of Science (JSPS) KAKENHI, Japan;
Consejo Nacional de Ciencia (CONACYT) y Tecnolog\'{i}a, through Fondo de Cooperaci\'{o}n Internacional en Ciencia y Tecnolog\'{i}a (FONCICYT) and Direcci\'{o}n General de Asuntos del Personal Academico (DGAPA), Mexico;
Nederlandse Organisatie voor Wetenschappelijk Onderzoek (NWO), Netherlands;
The Research Council of Norway, Norway;
Pontificia Universidad Cat\'{o}lica del Per\'{u}, Peru;
Ministry of Science and Higher Education, National Science Centre and WUT ID-UB, Poland;
Korea Institute of Science and Technology Information and National Research Foundation of Korea (NRF), Republic of Korea;
Ministry of Education and Scientific Research, Institute of Atomic Physics, Ministry of Research and Innovation and Institute of Atomic Physics and Universitatea Nationala de Stiinta si Tehnologie Politehnica Bucuresti, Romania;
Ministerstvo skolstva, vyskumu, vyvoja a mladeze SR, Slovakia;
National Research Foundation of South Africa, South Africa;
Swedish Research Council (VR) and Knut \& Alice Wallenberg Foundation (KAW), Sweden;
European Organization for Nuclear Research, Switzerland;
Suranaree University of Technology (SUT), National Science and Technology Development Agency (NSTDA) and National Science, Research and Innovation Fund (NSRF via PMU-B B05F650021), Thailand;
Turkish Energy, Nuclear and Mineral Research Agency (TENMAK), Turkey;
National Academy of  Sciences of Ukraine, Ukraine;
Science and Technology Facilities Council (STFC), United Kingdom;
National Science Foundation of the United States of America (NSF) and United States Department of Energy, Office of Nuclear Physics (DOE NP), United States of America.
In addition, individual groups or members have received support from:
Czech Science Foundation (grant no. 23-07499S), Czech Republic;
FORTE project, reg.\ no.\ CZ.02.01.01/00/22\_008/0004632, Czech Republic, co-funded by the European Union, Czech Republic;
European Research Council (grant nos. 101220549, 950692), European Union;
Deutsche Forschungs Gemeinschaft (DFG, German Research Foundation) ``Neutrinos and Dark Matter in Astro- and Particle Physics'' (grant no. SFB 1258), Germany;
CONVECS project, CUP C97H23001700002 FESR 2021-2027 program, Italy.

\end{acknowledgement}

%%%%%%%% Bibliography
\bibliographystyle{utphys}   % Remember we use title in the biblio
\bibliography{bibliography}
%\input {bibliography.tex}

%%%%%%%%%%%%%%%%%%%%%%%%%%%%%%%%
% Appendices: yours (if any) + authorlist
%%%%%%%%%%%%%%%%%%%%%%%%%%%%%%%%
\newpage
\appendix

%%%%% Authorlist - please do not touch: handled by EB chairs
\section{The ALICE Collaboration}
\label{app:collab}
% ALICE Collaboration author list for 2026-03-04
\begin{flushleft} 
\small

D.A.H.~Abdallah\,\orcidlink{0000-0003-4768-2718}\,$^{\rm 134}$, 
I.J.~Abualrob\,\orcidlink{0009-0005-3519-5631}\,$^{\rm 111}$, 
S.~Acharya\,\orcidlink{0000-0002-9213-5329}\,$^{\rm 49}$, 
K.~Agarwal\,\orcidlink{0000-0001-5781-3393}\,$^{\rm II,}$$^{\rm 23}$, 
G.~Aglieri Rinella\,\orcidlink{0000-0002-9611-3696}\,$^{\rm 32}$, 
L.~Aglietta\,\orcidlink{0009-0003-0763-6802}\,$^{\rm 24}$, 
N.~Agrawal\,\orcidlink{0000-0003-0348-9836}\,$^{\rm 25}$, 
Z.~Ahammed\,\orcidlink{0000-0001-5241-7412}\,$^{\rm 132}$, 
S.~Ahmad\,\orcidlink{0000-0003-0497-5705}\,$^{\rm 15}$, 
I.~Ahuja\,\orcidlink{0000-0002-4417-1392}\,$^{\rm 36}$
% ...

\section*{Affiliation Notes}

$^{\rm I}$ Deceased\\
$^{\rm II}$ Also at: INFN Trieste, Trieste, Italy\\
$^{\rm III}$ Also at: Fondazione Bruno Kessler (FBK), Trento, Italy\\
$^{\rm IV}$ Also at: Czech Technical University in Prague, Prague, Czech Republic\\
$^{\rm V}$ Also at: Instituto de Fisica da Universidade de Sao Paulo\\
$^{\rm VI}$ Also at: Dipartimento DET del Politecnico di Torino, Turin, Italy\\
$^{\rm VII}$ Also at: University College of Dublin, Dublin, Ireland\\
$^{\rm VIII}$ Also at: Department of Applied Physics, Aligarh Muslim University, Aligarh, India\\
$^{\rm IX}$ Also at: Institute of Theoretical Physics, University of Wroclaw, Wroclaw, Poland\\
$^{\rm X}$ Also at: Facultad de Ciencias, Universidad Nacional Aut\'{o}noma de M\'{e}xico, Mexico City, Mexico\\

\section*{Collaboration Institutes}

$^{1}$ A.I. Alikhanyan National Science Laboratory (Yerevan Physics Institute) Foundation, Yerevan, Armenia\\
$^{2}$ AGH University of Krakow, Cracow, Poland\\
$^{3}$ Bogolyubov Institute for Theoretical Physics, National Academy of Sciences of Ukraine, Kyiv, Ukraine\\
$^{4}$ Bose Institute, Department of Physics  and Centre for Astroparticle Physics and Space Science (CAPSS), Kolkata, India\\
$^{5}$ California Polytechnic State University, San Luis Obispo, California, United States\\
$^{6}$ Central China Normal University, Wuhan, China\\
$^{7}$ Centro de Aplicaciones Tecnol\'{o}gicas y Desarrollo Nuclear (CEADEN), Havana, Cuba\\
$^{8}$ Centro de Investigaci\'{o}n y de Estudios Avanzados (CINVESTAV), Mexico City and M\'{e}rida, Mexico\\
$^{9}$ Chicago State University, Chicago, Illinois, United States\\
$^{10}$ China Nuclear Data Center, China Institute of Atomic Energy, Beijing, China\\
$^{11}$ China University of Geosciences, Wuhan, China\\
$^{12}$ Chungbuk National University, Cheongju, Republic of Korea\\
$^{13}$ Comenius University Bratislava, Faculty of Mathematics, Physics and Informatics, Bratislava, Slovak Republic\\
$^{14}$ Creighton University, Omaha, Nebraska, United States\\
$^{15}$ Department of Physics, Aligarh Muslim University, Aligarh, India\\
$^{16}$ Department of Physics, Pusan National University, Pusan, Republic of Korea\\
$^{17}$ Department of Physics, Sejong University, Seoul, Republic of Korea\\
$^{18}$ Department of Physics, University of California, Berkeley, California, United States\\
$^{19}$ Department of Physics, University of Oslo, Oslo, Norway\\
$^{20}$ Department of Physics and Technology, University of Bergen, Bergen, Norway\\
$^{21}$ Dipartimento di Fisica, Universit\`{a} di Pavia, Pavia, Italy\\
$^{22}$ Dipartimento di Fisica dell'Universit\`{a} and Sezione INFN, Cagliari, Italy\\
$^{23}$ Dipartimento di Fisica dell'Universit\`{a} and Sezione INFN, Trieste, Italy\\
$^{24}$ Dipartimento di Fisica dell'Universit\`{a} and Sezione INFN, Turin, Italy\\
$^{25}$ Dipartimento di Fisica e Astronomia dell'Universit\`{a} and Sezione INFN, Bologna, Italy\\
$^{26}$ Dipartimento di Fisica e Astronomia dell'Universit\`{a} and Sezione INFN, Catania, Italy\\
$^{27}$ Dipartimento di Fisica e Astronomia dell'Universit\`{a} and Sezione INFN, Padova, Italy\\
$^{28}$ Dipartimento di Fisica `E.R.~Caianiello' dell'Universit\`{a} and Gruppo Collegato INFN, Salerno, Italy\\
$^{29}$ Dipartimento DISAT del Politecnico and Sezione INFN, Turin, Italy\\
$^{30}$ Dipartimento di Scienze MIFT, Universit\`{a} di Messina, Messina, Italy\\
$^{31}$ Dipartimento Interateneo di Fisica `M.~Merlin' and Sezione INFN, Bari, Italy\\
$^{32}$ European Organization for Nuclear Research (CERN), Geneva, Switzerland\\
$^{33}$ Faculty of Electrical Engineering, Mechanical Engineering and Naval Architecture, University of Split, Split, Croatia\\
$^{34}$ Faculty of Nuclear Sciences and Physical Engineering, Czech Technical University in Prague, Prague, Czech Republic\\
$^{35}$ Faculty of Physics, Sofia University, Sofia, Bulgaria\\
$^{36}$ Faculty of Science, P.J.~\v{S}af\'{a}rik University, Ko\v{s}ice, Slovak Republic\\
$^{37}$ Faculty of Technology, Environmental and Social Sciences, Bergen, Norway\\
$^{38}$ Frankfurt Institute for Advanced Studies, Johann Wolfgang Goethe-Universit\"{a}t Frankfurt, Frankfurt, Germany\\
$^{39}$ Fudan University, Shanghai, China\\
$^{40}$ Gauhati University, Department of Physics, Guwahati, India\\
$^{41}$ Helmholtz-Institut f\"{u}r Strahlen- und Kernphysik, Rheinische Friedrich-Wilhelms-Universit\"{a}t Bonn, Bonn, Germany\\
$^{42}$ Helsinki Institute of Physics (HIP), Helsinki, Finland\\
$^{43}$ High Energy Physics Group,  Universidad Aut\'{o}noma de Puebla, Puebla, Mexico\\
$^{44}$ Horia Hulubei National Institute of Physics and Nuclear Engineering, Bucharest, Romania\\
$^{45}$ HUN-REN Wigner Research Centre for Physics, Budapest, Hungary\\
$^{46}$ Indian Institute of Technology Bombay (IIT), Mumbai, India\\
$^{47}$ Indian Institute of Technology Indore, Indore, India\\
$^{48}$ INFN, Laboratori Nazionali di Frascati, Frascati, Italy\\
$^{49}$ INFN, Sezione di Bari, Bari, Italy\\
$^{50}$ INFN, Sezione di Bologna, Bologna, Italy\\
$^{51}$ INFN, Sezione di Cagliari, Cagliari, Italy\\
$^{52}$ INFN, Sezione di Catania, Catania, Italy\\
$^{53}$ INFN, Sezione di Padova, Padova, Italy\\
$^{54}$ INFN, Sezione di Pavia, Pavia, Italy\\
$^{55}$ INFN, Sezione di Torino, Turin, Italy\\
$^{56}$ INFN, Sezione di Trieste, Trieste, Italy\\
$^{57}$ Inha University, Incheon, Republic of Korea\\
$^{58}$ Institute for Gravitational and Subatomic Physics (GRASP), Utrecht University/Nikhef, Utrecht, Netherlands\\
$^{59}$ Institute of Experimental Physics, Slovak Academy of Sciences, Ko\v{s}ice, Slovak Republic\\
$^{60}$ Institute of Physics, Homi Bhabha National Institute, Bhubaneswar, India\\
$^{61}$ Institute of Physics of the Czech Academy of Sciences, Prague, Czech Republic\\
$^{62}$ Institute of Space Science (ISS), Bucharest, Romania\\
$^{63}$ Institut f\"{u}r Kernphysik, Johann Wolfgang Goethe-Universit\"{a}t Frankfurt, Frankfurt, Germany\\
$^{64}$ Instituto de Ciencias Nucleares, Universidad Nacional Aut\'{o}noma de M\'{e}xico, Mexico City, Mexico\\
$^{65}$ Instituto de F\'{i}sica, Universidade Federal do Rio Grande do Sul (UFRGS), Porto Alegre, Brazil\\
$^{66}$ Instituto de F\'{\i}sica, Universidad Nacional Aut\'{o}noma de M\'{e}xico, Mexico City, Mexico\\
$^{67}$ iThemba LABS, National Research Foundation, Somerset West, South Africa\\
$^{68}$ Jeonbuk National University, Jeonju, Republic of Korea\\
$^{69}$ Korea Institute of Science and Technology Information, Daejeon, Republic of Korea\\
$^{70}$ Laboratoire de Physique Subatomique et de Cosmologie, Universit\'{e} Grenoble-Alpes, CNRS-IN2P3, Grenoble, France\\
$^{71}$ Lawrence Berkeley National Laboratory, Berkeley, California, United States\\
$^{72}$ Lund University Department of Physics, Division of Particle Physics, Lund, Sweden\\
$^{73}$ Marietta Blau Institute, Vienna, Austria\\
$^{74}$ Nagasaki Institute of Applied Science, Nagasaki, Japan\\
$^{75}$ Nara Women{'}s University (NWU), Nara, Japan\\
$^{76}$ National Centre for Nuclear Research, Warsaw, Poland\\
$^{77}$ National Institute of Science Education and Research, Homi Bhabha National Institute, Jatni, India\\
$^{78}$ National Nuclear Research Center, Baku, Azerbaijan\\
$^{79}$ National Research and Innovation Agency - BRIN, Jakarta, Indonesia\\
$^{80}$ Niels Bohr Institute, University of Copenhagen, Copenhagen, Denmark\\
$^{81}$ Nikhef, National institute for subatomic physics, Amsterdam, Netherlands\\
$^{82}$ Nuclear Physics Group, STFC Daresbury Laboratory, Daresbury, United Kingdom\\
$^{83}$ Nuclear Physics Institute of the Czech Academy of Sciences, Husinec-\v{R}e\v{z}, Czech Republic\\
$^{84}$ Oak Ridge National Laboratory, Oak Ridge, Tennessee, United States\\
$^{85}$ Ohio State University, Columbus, Ohio, United States\\
$^{86}$ Physics Department, Panjab University, Chandigarh, India\\
$^{87}$ Physics Department, University of Jammu, Jammu, India\\
$^{88}$ Physics Program and International Institute for Sustainability with Knotted Chiral Meta Matter (WPI-SKCM$^{2}$), Hiroshima University, Hiroshima, Japan\\
$^{89}$ Physikalisches Institut, Eberhard-Karls-Universit\"{a}t T\"{u}bingen, T\"{u}bingen, Germany\\
$^{90}$ Physikalisches Institut, Ruprecht-Karls-Universit\"{a}t Heidelberg, Heidelberg, Germany\\
$^{91}$ Physik Department, Technische Universit\"{a}t M\"{u}nchen, Munich, Germany\\
$^{92}$ Politecnico di Bari and Sezione INFN, Bari, Italy\\
$^{93}$ Research Division and ExtreMe Matter Institute EMMI, GSI Helmholtzzentrum f\"ur Schwerionenforschung GmbH, Darmstadt, Germany\\
$^{94}$ Saga University, Saga, Japan\\
$^{95}$ Saha Institute of Nuclear Physics, Homi Bhabha National Institute, Kolkata, India\\
$^{96}$ School of Physics and Astronomy, University of Birmingham, Birmingham, United Kingdom\\
$^{97}$ Secci\'{o}n F\'{\i}sica, Departamento de Ciencias, Pontificia Universidad Cat\'{o}lica del Per\'{u}, Lima, Peru\\
$^{98}$ SUBATECH, IMT Atlantique, Nantes Universit\'{e}, CNRS-IN2P3, Nantes, France\\
$^{99}$ Sungkyunkwan University, Suwon City, Republic of Korea\\
$^{100}$ Suranaree University of Technology, Nakhon Ratchasima, Thailand\\
$^{101}$ Technical University of Ko\v{s}ice, Ko\v{s}ice, Slovak Republic\\
$^{102}$ The Henryk Niewodniczanski Institute of Nuclear Physics, Polish Academy of Sciences, Cracow, Poland\\
$^{103}$ The University of Texas at Austin, Austin, Texas, United States\\
$^{104}$ Universidad Aut\'{o}noma de Sinaloa, Culiac\'{a}n, Mexico\\
$^{105}$ Universidade de S\~{a}o Paulo (USP), S\~{a}o Paulo, Brazil\\
$^{106}$ Universidade Estadual de Campinas (UNICAMP), Campinas, Brazil\\
$^{107}$ Universidade Federal do ABC, Santo Andre, Brazil\\
$^{108}$ Universitatea Nationala de Stiinta si Tehnologie Politehnica Bucuresti, Bucharest, Romania\\
$^{109}$ University of Cape Town, Cape Town, South Africa\\
$^{110}$ University of Derby, Derby, United Kingdom\\
$^{111}$ University of Houston, Houston, Texas, United States\\
$^{112}$ University of Jyv\"{a}skyl\"{a}, Jyv\"{a}skyl\"{a}, Finland\\
$^{113}$ University of Kansas, Lawrence, Kansas, United States\\
$^{114}$ University of Liverpool, Liverpool, United Kingdom\\
$^{115}$ University of Science and Technology of China, Hefei, China\\
$^{116}$ University of Silesia in Katowice, Katowice, Poland\\
$^{117}$ University of South-Eastern Norway, Kongsberg, Norway\\
$^{118}$ University of Tennessee, Knoxville, Tennessee, United States\\
$^{119}$ University of the Witwatersrand, Johannesburg, South Africa\\
$^{120}$ University of Tokyo, Tokyo, Japan\\
$^{121}$ University of Tsukuba, Tsukuba, Japan\\
$^{122}$ University of Zagreb Faculty of Science, Department of Physics, Zagreb, Croatia\\
$^{123}$ Universit\"{a}t M\"{u}nster, Institut f\"{u}r Kernphysik, M\"{u}nster, Germany\\
$^{124}$ Universit\'{e} Clermont Auvergne, CNRS/IN2P3, LPC, Clermont-Ferrand, France\\
$^{125}$ Universit\'{e} de Lyon, CNRS/IN2P3, Institut de Physique des 2 Infinis de Lyon, Lyon, France\\
$^{126}$ Universit\'{e} de Strasbourg, CNRS, IPHC UMR 7178, F-67000 Strasbourg, France, Strasbourg, France\\
$^{127}$ Universit\'{e} Paris-Saclay, Centre d'Etudes de Saclay (CEA), IRFU, D\'{e}partment de Physique Nucl\'{e}aire (DPhN), Saclay, France\\
$^{128}$ Universit\'{e}  Paris-Saclay, CNRS/IN2P3, IJCLab, Orsay, France\\
$^{129}$ Universit\`{a} degli Studi di Foggia, Foggia, Italy\\
$^{130}$ Universit\`{a} del Piemonte Orientale, Vercelli, Italy\\
$^{131}$ Universit\`{a} di Brescia, Brescia, Italy\\
$^{132}$ Variable Energy Cyclotron Centre, Homi Bhabha National Institute, Kolkata, India\\
$^{133}$ Warsaw University of Technology, Warsaw, Poland\\
$^{134}$ Wayne State University, Detroit, Michigan, United States\\
$^{135}$ Yale University, New Haven, Connecticut, United States\\
$^{136}$ Yildiz Technical University, Istanbul, Turkey\\
$^{137}$ Yonsei University, Seoul, Republic of Korea\\
$^{138}$ Affiliated with an institute formerly covered by a cooperation agreement with CERN\\
$^{139}$ Affiliated with an international laboratory covered by a cooperation agreement with CERN.\\

\end{flushleft} 

\end{document}